\documentclass[]{pasj02} 
\usepackage[switch,mathlines]{lineno} 

\jyear{2026}
\Received{}
\Accepted{}

\begin{document} 
\nolinenumbers

\title{Autoencoder-based framework for anomaly detection in stellar spectra: application to the MaNGA Stellar Library}


\author{
 Akihiro \textsc{Suzuki}\altaffilmark{1}\altemailmark\orcid{0000-0002-7043-6112} \email{akihiro.suzuki@resceu.s.u-tokyo.ac.jp} 
}

\altaffiltext{1}{Research Center for the Early Universe, The University of Tokyo, 7-3-1 Hongo, Bunkyo-ku, Tokyo 181-8588, Japa}



\KeyWords{methods: data analysis --- stars: general --- stars: carbon --- stars: chemically peculiar}  

\maketitle

\begin{abstract}
A machine-learning–based method is developed to identify objects with unusual stellar spectra. 
The method employs an autoencoder, a neural network trained to compress spectral data into a low-dimensional representation and subsequently reconstruct it. 
Spectra that deviate significantly from the dominant patterns in the training dataset are identified using the reconstruction error as an anomaly score. 
The models are applied to selected datasets from the MaNGA Stellar Library, an empirical library of stellar spectra. 
Several spectra are flagged as anomalous: an object with likely instrumental and/or reduction issues, two carbon stars, and an oxygen-rich thermally pulsating asymptotic giant branch star. 
The sources of the large reconstruction errors are examined, and the effectiveness and limitations of autoencoder-based approaches for detecting anomalous stellar spectra are discussed.
\end{abstract}


\section{Introduction}\label{sec:introduction} 
Stellar spectral classification has long played a central role in astrophysics, providing a physically meaningful framework for describing stellar atmospheres and evolutionary states (e.g., \cite{1943assw.book.....M,1983ApJS...52..121G,1984ApJS...56..257J,2009ssc..book.....G,2011A&A...532A..95F}). By organizing stars according to their spectral features, such as absorption line strengths and continua shaped by temperature, surface gravity, and chemical composition, classification schemes have enabled systematic studies of stellar populations. 

Large stellar spectroscopic surveys are now a foundation of modern astrophysics, supporting detailed studies of stellar populations and evolution, as well as Galactic structure. 
Surveys such as SDSS (\cite{2000AJ....120.1579Y}), LAMOST (\cite{2012RAA....12.1197C}), APOGEE (\cite{2017AJ....154...94M}), GALAH (\cite{2015MNRAS.449.2604D,2017MNRAS.465.3203M}), and Gaia-ESO (\cite{2012Msngr.147...25G,2013Msngr.154...47R}) have produced millions of spectra with broad wavelength coverage and varying resolutions. 
While these datasets provide unprecedented statistical power, their size and complexity pose challenges for traditional analysis methods. 
Therefore, robust and scalable methods for spectral analysis have become increasingly essential. 
This situation motivates the development of automated, data-driven approaches capable of capturing both common patterns and rare or anomalous objects (e.g., \cite{1994ApJ...426..340G,1998MNRAS.298..361B,1998MNRAS.295..312S,2020MNRAS.491.2280S}).

Anomaly detection in stellar spectra is of particular interest, as outliers may correspond to astrophysically meaningful objects such as chemically peculiar stars, rare evolutionary stages, interacting binaries, emission-line objects, or data quality issues. 
Classical approaches to anomaly detection often rely on handcrafted features, physical model fitting, or distance-based metrics in low-dimensional parameter spaces, such as color-color diagrams. 
However, these techniques may struggle to scale to large datasets and perhaps more importantly, they may be biased toward well-known spectral classes, limiting their ability to uncover genuinely novel or unexpected objects/phenomena.

In recent years, unsupervised and self-supervised machine learning methods have emerged as powerful tools for representation learning in high-dimensional data and have been applied to a wide range of problems in astronomy (e.g., \cite{2015MNRAS.452.4183H,2017MNRAS.465.4530B,2019MNRAS.484..834G,2019MNRAS.487.2874I,2019MNRAS.489.3591P,2021A&C....3600481L,2021MNRAS.508.2946S}). 
Unsupervised approaches to identifying anomalous stellar spectra have previously been explored in the literature. 
For example, \citet{2013MNRAS.431.1800W} combined statistical and dimensionality-reduction techniques for identifying objects with peculiar spectra. 
Reis et al.~(2018) developed a method based on unsupervised random forest to search for outliers in APOGEE samples. 

Autoencoder-based approaches (e.g., \cite{Hinton2006,Chalapathy2017}) are particularly well suited to astronomical spectra: they naturally handle high-dimensional spectral data, can model nonlinear correlations between spectral features, and are flexible enough to accommodate varying signal-to-noise ratios and instrumental effects. Moreover, the learned latent space can provide additional insight into the structure of the dataset, potentially revealing continuous trends or clusters related to stellar parameters such as effective temperature, metallicity, and surface gravity.
Owing to their broad applicability, autoencoders and related techniques are increasingly used for stellar spectral datasets from large spectroscopic surveys. 
Previous examples include the automatic estimation of stellar atmospheric parameters (\cite{2015MNRAS.452..158Y}), the detection of emission-line stars in GALAH data (\cite{2021MNRAS.500.4849C}), and the identification of emission features associated with stellar magnetic activity in LAMOST data (\cite{2022MNRAS.514.4781X}), and so on. 

Rapidly increasing applications of autoencoder-based anomaly detection to astronomical spectra from large spectroscopic datasets (e.g., \cite{2023MNRAS.526.3072B}; \cite{2023ApJ...956L...6L}; \cite{2025A&A...703A.242O}; \cite{2025MNRAS.543..691Z}) demonstrate the effectiveness of such methods in capturing the intrinsic structure of spectroscopic data. 
However, most of these studies focus on galaxy spectra, whose diversity and physical drivers differ significantly from those of stellar atmospheres. 
In contrast, the application of reconstruction-based deep learning methods to homogeneous stellar spectral libraries remains comparatively less explored.

Unlike galaxy spectra, which are composed of composite stellar populations, active galactic nuclei, and multiple interstellar components, stellar spectra are governed primarily by a small set of atmospheric parameters, including effective temperature, surface gravity, and chemical composition. 
As a result, deviations from the dominant population may arise from a variety of distinct causes, including astrophysical peculiarities (e.g., chemically unusual stars), sparsely sampled regions of parameter space, unresolved composite systems, or data reduction artifacts. 
A meaningful anomaly-detection framework must therefore not only identify spectra with large reconstruction residuals, but also clarify the physical or instrumental origins of those deviations. 
Systematically characterizing the patterns of anomalies in homogeneous stellar spectral libraries is thus an essential step toward interpreting unsupervised outlier detection in an astrophysical context.

In this work, I investigate the application of autoencoder-based anomaly detection to the MaNGA Stellar Library (\cite{2019ApJ...883..175Y}), an empirical stellar spectral library collected by Mapping Nearby Galaxies
at Apache Point Observatory (MaNGA; \cite{2015ApJ...798....7B}) and assembled as part of the SDSS-IV survey (\cite{2017AJ....154...28B}). 
Unsupervised autoencoder models are trained on a representative subset of spectra and reconstruction error is used as an anomaly score to identify anomalous objects. 
Then, the nature of the detected anomalies is analyzed for assessing their astrophysical relevance, demonstrating that this approach provides an effective, data-driven framework for discovering rare or unusual stellar spectra in modern survey datasets. 

This paper is structured as follows. In Section \ref{sec:data}, the datasets used in this study are described. 
In Section \ref{sec:neural_net}, the autoencoder architecture and training strategy are described, and the trained models are applied to the selected dataset. 
In Section \ref{sec:unsuccess}, spectra that are unsuccessfully reconstructed are presented and we discuss the origins of the large reconstruction errors. 
Finally, Section \ref{sec:summary} concludes this paper. 

\section{Training data and preprocessing}\label{sec:data}
\begin{figure}
\begin{center}
\includegraphics[scale=0.75]{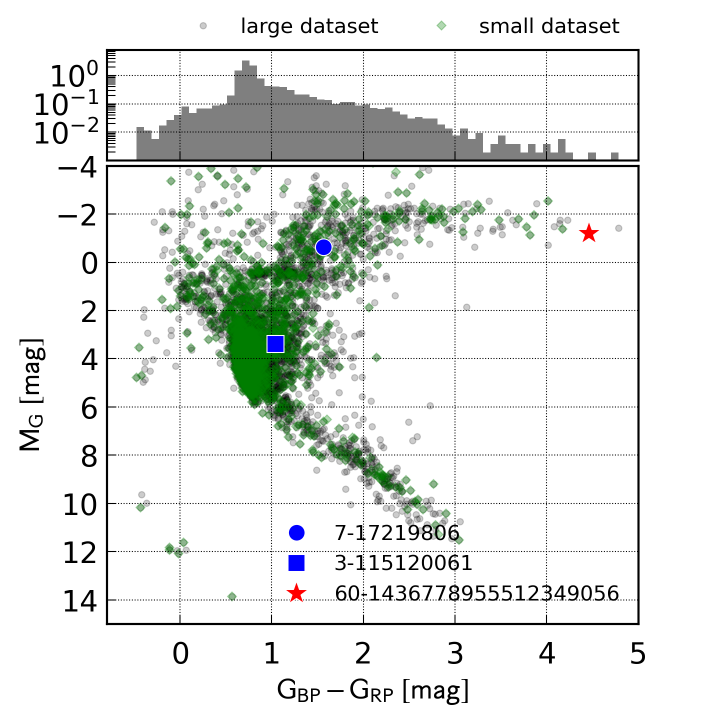}
\end{center}
\caption{Color-magnitude diagram (absolute $G$-band magnitude vs BP-RP color) for the samples used in this study. 
The samples from \emph{small} and \emph{large datasets} are plotted as small diamonds and circles. 
Three objects flagged as outliers by the autoencoder models (see Section \ref{sec:unsuccess}) are also plotted as a big circle, square, and star as indicated in the lower panel. 
In the upper panel, the distribution of the BP-RP color for the \emph{large dataset} is also presented as a histogram. 
{Alt text: Scatter plot showing the distribution of sample data and a hitogram on top of it. }}
\label{fig:cmd}
\end{figure}

\begin{figure*}
\begin{center}
\includegraphics[scale=0.8]{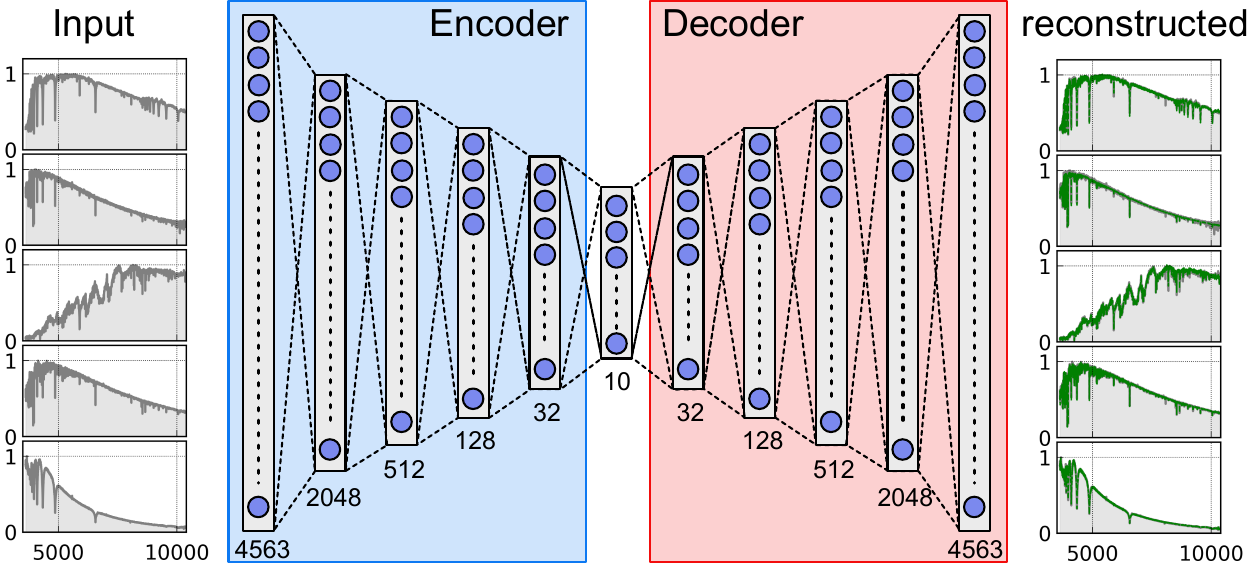}
\end{center}
\caption{Schematic representation of the autoencoder model in this work. 
{Alt text: Schematic illustration showing the model. }
}
\label{fig:schematic}
\end{figure*}

\subsection{Data selection}
The datasets are constructed using spectra from the MaNGA Stellar Library (MaStar), which provides flux-calibrated optical spectra obtained with the MaNGA integral-field spectrographs (\cite{2015AJ....149...77D}), covering a wavelength range of approximately $3622$–$10354\,\mathring{\mathrm{A}}$ at a resolving power of $R \sim 1800$.
The latest MaStar data are distributed as part of SDSS Data Release 17 \citep{2022ApJS..259...35A}. 
With the goal of covering as wide parameter space (effective temperature, surface gravity, composition, and so on) as possible (\cite{2020ApJ...899...62C,2020MNRAS.496.2962M,2022A&A...668A..21L,2022MNRAS.509.4308H,2022MNRAS.517.4275H,2022AJ....163...56I}), the data products are not intended to provide unbiased and complete statistical samples. 
Nevertheless, it is one of the largest stellar spectral dataset maintaining homogeneous data quality and used for classifiation by supervized machine-larning techniques (e.g., \cite{2025A&A...693A.300E}) and identifying stellar spectra with unnusual features, e.g., thermally pulsating asymptotic giant branch (TP-AGB) stars (\cite{2024MNRAS.530.1534H}). 
In this work, the latest value-added catalogs and spectral data products\footnote{available at https://www.sdss4.org/dr17/mastar/} are used. 

MaStar delivers stellar spectra with well-controlled and homogeneous data quality.
This study adopts the MaStar Uniform Resolution Spectra, which are convolved to a common wavelength-dependent line-spread function corresponding to the 60th and 90th percentiles of the MaStar resolution distribution (see, \cite{2019ApJ...883..175Y}, for details).
Prior to sample construction, spectra containing zero flux values in any wavelength bin are excluded to ensure well-defined inputs across the full wavelength range.
This selection yields 2770 and 6522 stellar spectra from the parent datasets corresponding to the 60th and 90th percentile resolutions, respectively.
Hereafter, these are referred to as the \emph{small} and \emph{large datasets}.

The \emph{small dataset} is used for training and validation, while the trained models are applied to the \emph{large dataset}.
From the \emph{small dataset}, ten independent training realizations are generated by randomly selecting 2000 spectra in each case; the remaining 770 spectra are used for validation.
These multiple random subsets allow us to assess the robustness of the autoencoder training against sample variance while preserving the stellar spectral diversity and a homogeneous instrumental resolution. 

No explicit filtering is applied based on stellar classification or atmospheric parameters, allowing the model to learn directly from the data distribution.
Objects with potential artificial issues, as well as rare spectral types, are therefore retained and are expected to appear as outliers in the learned representation.
This choice enables a fully data-driven search for anomalous spectra without imposing strong astrophysical priors.

Since the MaStar samples are drawn from multiple photometric surveys, a cross-matched catalog is also provided.
In particular, the information from Gaia mission (\cite{2016A&A...595A...1G,2016A&A...595A...2G,2018A&A...616A...1G,2023A&A...674A...1G}), including Gaia ID and information, absolute G-band magnitude ($\mathrm{M}_G$), and the BP–RP color, is available along with distance estimates from \citet{2021AJ....161..147B}.
This information is used to characterize the properties of objects identified by the anomaly detection method.
Out of the 2770 and 6522 objects in the \emph{small} and \emph{large datasets}, 2765 and 6501 objects have corresponding Gaia IDs.
Figure~\ref{fig:cmd} shows the color–absolute magnitude diagram ($\mathrm{M}_G$ vs $\mathrm{G}_{BP}$-$\mathrm{G}_{RP}$) for the \emph{small} and \emph{large datasets}.
The samples include a large number of main-sequence stars, extending to low-mass dwarfs, as well as evolved stars in the red clump and asymptotic giant branch, and a small population of white dwarfs.

\subsection{Preprocessing}

Prior to training, each spectrum is normalized by its maximum flux value to remove overall flux scaling.
Each spectrum consists of 4563 wavelength bins.
The normalized flux can therefore be represented as a 4563-dimensional vector ($N_\mathrm{dim}=4563$), hereafter referred to as the flux vector, $F_i$ $(i = 1,\ldots,N_\mathrm{dim})$, or simply denoted by $\mathbf{F}$.
The logarithm of the normalized flux is then used as the input feature vector for the autoencoder, such that $f_i = \ln(F_i)$ or $\mathbf{f}=\ln(\mathbf{F})$.
Because spectra containing zero flux values are excluded during the data selection above, the logarithmic transformation is well defined for all wavelength bins.
This preprocessing step reduces the dynamic range of the input data, allowing the model to focus on relative spectral variations rather than absolute flux differences.

\section{Autoencoder architecture, training, and application}\label{sec:neural_net}
In this section, the architecture of the autoencoder models, the training strategy, and the application of the trained models are described. 

\subsection{Network architecture}
\begin{figure}
\begin{center}
\includegraphics[scale=0.75]{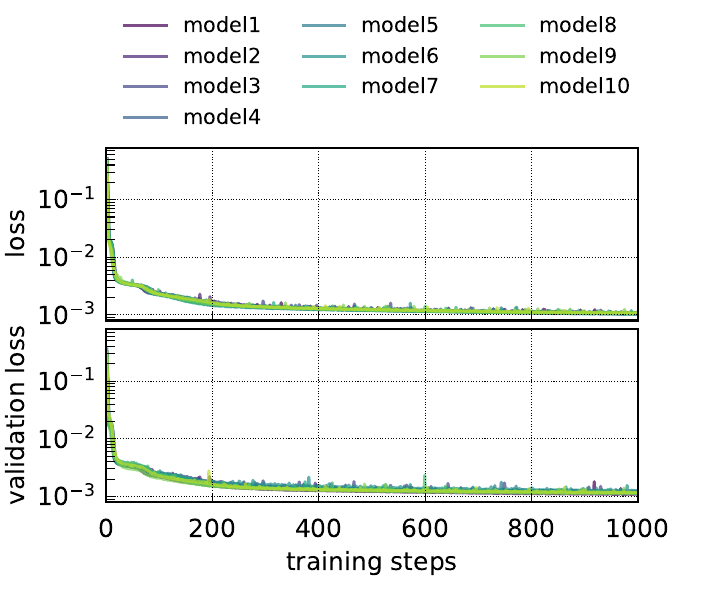}
\end{center}
\caption{Training (upper) and validation (lower) losses as a function of the training epochs. In each panel, all the ten autoencoder models are presented. 
{Alt text: Tow plots with ten lines each. }}
\label{fig:loss}
\end{figure}

Figure~\ref{fig:schematic} presents a schematic representation of the autoencoder model trained and used in this work. Autoencoders are a class of neural networks designed to learn compact representations of high-dimensional data by compressing the input into a lower-dimensional latent space and then reconstructing it. More detailed introductions to neural networks and autoencoders can be found elsewhere (e.g., \cite{BN06,2019arXiv190407248B,2025arXiv251010713T}). 
The following is a brief and intuitive description of the architecture. 

A feedforward neural network consists of a sequence of layers that progressively transform the input data. Each layer performs a linear transformation followed by a nonlinear operation, allowing the network to model complex relationships between the input and output. 
Mathematically, the transformation at layer $l$ is given by
\begin{equation}
\mathbf{h}^{(l)} = \sigma\big(\mathbf{W}^{(l)} \mathbf{h}^{(l-1)} + \mathbf{b}^{(l)}\big),
\end{equation}
where $\mathbf{h}^{(l-1)}$ and $\mathbf{h}^{(l)}$ denote the input and output vectors of layer $l$, $\mathbf{W}^{(l)}$ and $\mathbf{b}^{(l)}$ are trainable weight matrices and bias vectors, and $\sigma$ is a nonlinear activation function. 
By stacking multiple layers of linear transformations and nonlinear activations, the network can approximate highly nonlinear mappings from the input space to the output space. 
In the present application, the input for the initial layer is set to the transformed flux vector; $\mathbf{h}^{(0)}=\mathbf{f}$, while the final layer returns the reconstructed output $\hat{\mathbf{f}}$ with the same dimension as the input $\mathbf{f}$. 
The network parameters $\mathbf{W}^{(l)}$ and $\mathbf{b}^{(l)}$ are learned during training by iteratively adjusting them to minimize a loss function that quantifies the discrepancy between the input vector $\mathbf{f}$ and the output $\hat{\mathbf{f}}$ (see, Sec.~\ref{sec:results}). 
This optimization is in general performed using gradient-based methods.

The proposed model is a symmetric, fully connected autoencoder composed of an encoder and a decoder (Figure~\ref{fig:schematic}).
The encoder maps the 4563-dimensional input flux vector through a sequence of dense layers with decreasing dimensionality (2048, 512, 128, and 32) to a 10-dimensional latent representation.
This bottleneck layer enforces a compact encoding of the input features.
The decoder mirrors the encoder architecture, reconstructing the input by progressively increasing the dimensionality through dense layers of sizes 32, 128, 512, and 2048, followed by an output layer of size 4563.
Rectified linear unit (ReLU) activations (\cite{NH10,GBB11}), defined as $\sigma(x) = \max(0,x)$, are used in all hidden layers, while a linear activation is applied at the output layer.

The specific choice of network depth and layer widths is not unique and was not determined through an exhaustive hyperparameter search.
Instead, the adopted configuration represents a practical compromise between model capacity and computational efficiency, designed to capture the dominant structure of the input spectra while maintaining a compact latent representation.
As is common in autoencoder-based approaches, alternative architectural choices may yield comparable performance.
Nevertheless, as demonstrated below, the selected architecture achieves stable training and provides accurate reconstructions of the input data, indicating that it is sufficiently expressive for the purposes of this study.

\begin{figure*}
\begin{center}
\includegraphics[scale=0.90]{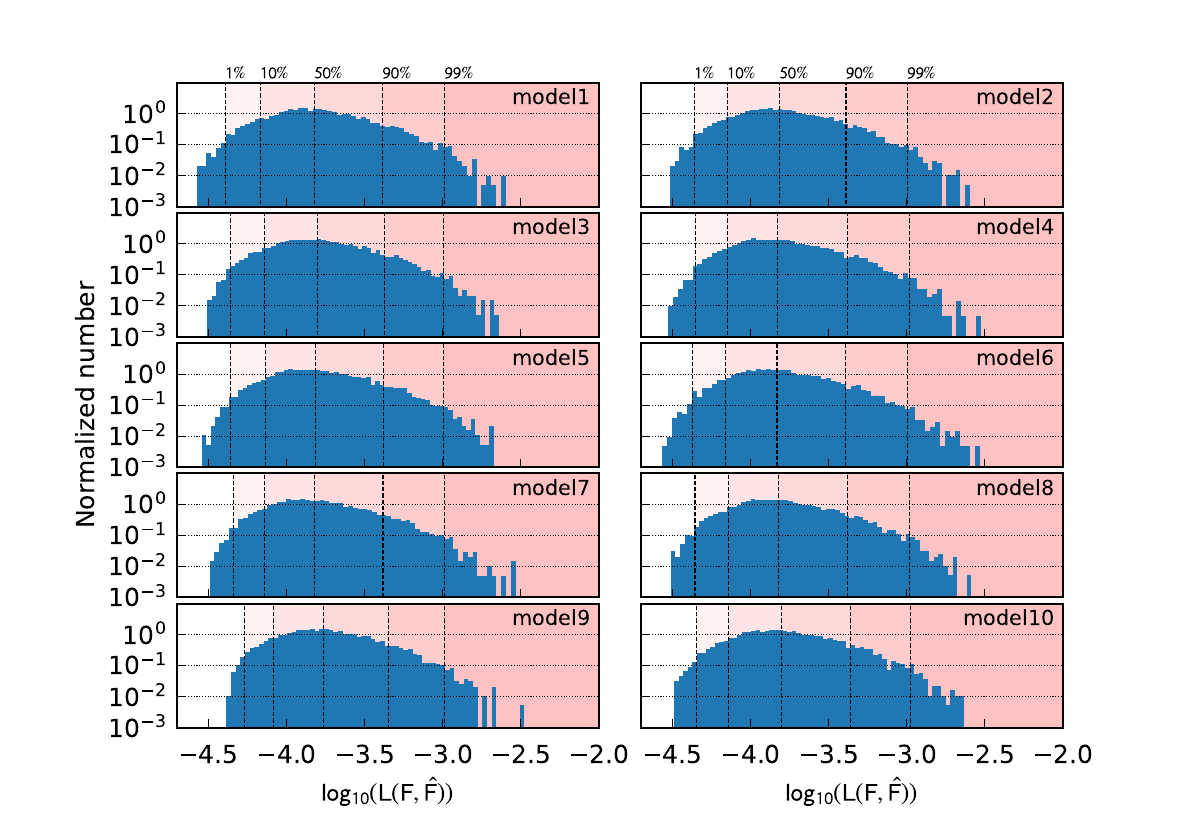}
\end{center}
\caption{Reconstruction error distributions for all the ten models. In each panel, the distribution is plotted as a histogram. 
The 1st, 10th, 50th, 90th, and 99th percentiles of the distributions are indicated by vertical dashes lines (from left to right). 
{Alt text: ten histograms showing distribution of errors in ten different models. }}
\label{fig:error}
\end{figure*}

\subsection{Implementation and training}\label{sec:results}
The network is implemented and trained using TensorFlow \citep{TensorFlow2016} with the Keras API \citep{Keras2015}, which provides a high-level interface for defining and training deep neural networks.
Training was performed using the Adam (Adaptive Moment Estimation) optimizer (\cite{2014arXiv:1412.6980}) with a fixed learning rate of 0.001 and a batch size of 32.
The autoencoder is trained to minimize the reconstruction error between the input (transformed) flux vector $\mathbf{f}$ and its reconstructed output $\hat{\mathbf{f}}$.
The mean squared error (MSE) is employed,
\begin{equation}
\mathcal{L}(\mathbf{f}, \hat{\mathbf{f}}) =
\frac{1}{N_\mathrm{dim}} \sum_{i=1}^{N_\mathrm{dim}} (f_i - \hat{f}_i)^2,
\label{eq:mse}
\end{equation}
as the loss function. 
The training loss is computed as the average reconstruction error over the training dataset and serves as a proxy for the model's ability to recover the salient spectral features of the training data.

To assess the generalization capability of the model and to prevent overfitting, a validation loss is computed using the same reconstruction error metric.
As shown below, only a small discrepancy is observed between the training and validation losses, indicating that the model generalizes well and does not suffer from significant overfitting. 
Using the ten realized training and validation data (created from the \emph{small dataset}) described above, ten autoencoder models with identical architectures and hyperparameters are  independently trained. 
Hereafter they are referred to as {\sc model}1 through {\sc model}10. 
After training, the learned models are applied to the \emph{large dataset}.

Figure~\ref{fig:loss} shows the training and validation losses as a function of training epochs.
Both losses exhibit similar convergence behavior.
The losses rapidly decrease to $\sim 3 \times 10^{-3}$ within the first several tens of epochs and then gradually decrease to $\sim 10^{-3}$, with only small temporal increases.
This trend is consistently reproduced across all ten models, demonstrating the robustness of the training process.

\subsection{Application to the large sample}

\begin{figure*}
\begin{center}
\includegraphics[scale=0.70]{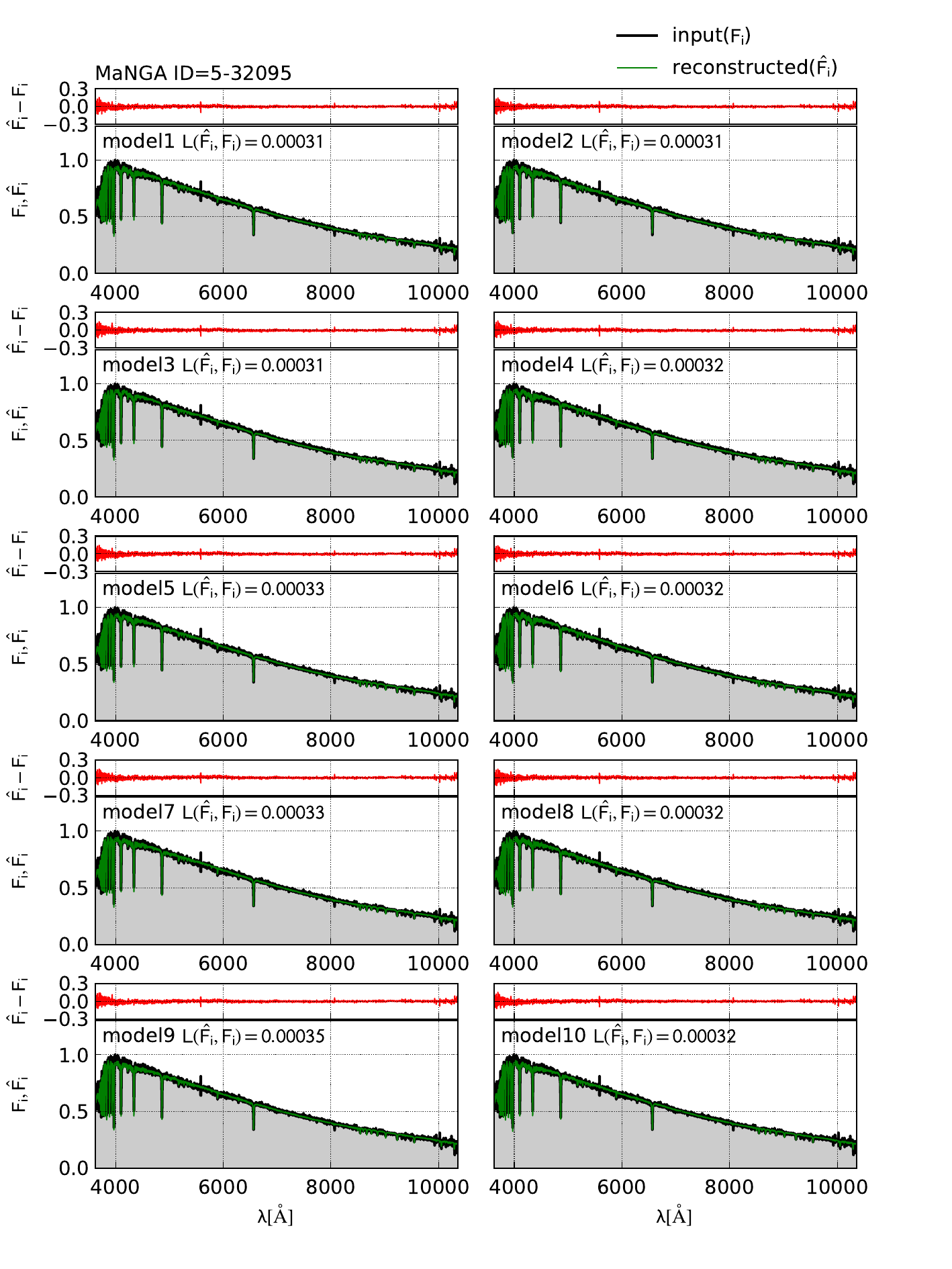}
\end{center}
\caption{Resoncstructed stellar spectra for a randomly selected object from the large sample. 
{Alt text: Ten plots showing the model and observed spectra wit thin and shick lines and residuapl plotted on top of each panel. }}
\label{fig:A_star}
\end{figure*}

The autoencoder models trained on the \emph{small dataset} are applied to the \emph{large dataset}.
For each object in the \emph{large dataset}, the input spectrum is reconstructed and the reconstruction error is computed independently using each of the ten trained autoencoder models.
The reconstruction error is defined as the mean squared error (Eq.~\ref{eq:mse}), evaluated for the normalized flux vectors $\mathbf{F}$ and $\hat{\mathbf{F}}$, ${\cal L}(\mathbf{F},\hat{\mathbf{F}})$,
where the reconstructed flux vector $\hat{\mathbf{F}}$ is obtained as $\hat{F_i} = \exp(\hat{f_i})$. 
Objects exhibiting large reconstruction errors consistently across multiple models are flagged as potential anomalies.
This ensemble-based approach mitigates sensitivity to individual model initializations and provides a more robust identification of anomalous spectra.

Figure~\ref{fig:error} shows the distributions of reconstruction errors for the 6522 spectra in the \emph{large dataset}.
All ten autoencoder models produce similar error distributions, spanning approximately $\log_{10} \mathcal{L}(\mathbf{F}, \hat{\mathbf{F}}) \simeq -4.5$ to $-2.5$.
The 1st, 10th, 50th, 90th, and 99th percentiles of the distribution are located at $\log_{10} \mathcal{L}(\mathbf{F}, \hat{\mathbf{F}})\simeq -4.4$, $-4.2$, $-3.8$, $-3.4$, and $-3.0$, respectively, with exact values listed in Table~\ref{table:error_percentiles}.

\begin{table}
\caption{Properties of error distributions presented in Figure~\ref{fig:error}}
\centering
\label{table:error_percentiles}
\begin{tabular}{r|rrrrr}
model&\multicolumn{5}{c}{percentile}\\
&$1\%$&$10\%$&$50\%$&$90\%$&$99\%$\\
\hline\hline
{\sc model1}&$-4.391$&$ -4.169$&$ -3.823$&$ -3.386$&$ -2.989$\\
{\sc model2}&$-4.360$&$ -4.150$&$ -3.815$&$ -3.389$&$ -2.997$\\
{\sc model3}&$-4.356$&$ -4.141$&$ -3.801$&$ -3.370$&$ -2.994$\\
{\sc model4}&$-4.352$&$ -4.146$&$ -3.825$&$ -3.379$&$ -2.985$\\
{\sc model5}&$-4.356$&$ -4.135$&$ -3.813$&$ -3.377$&$ -2.993$\\
{\sc model6}&$-4.373$&$ -4.161$&$ -3.831$&$ -3.392$&$ -2.998$\\
{\sc model7}&$-4.342$&$ -4.140$&$ -3.821$&$ -3.383$&$ -2.988$\\
{\sc model8}&$-4.355$&$ -4.142$&$ -3.818$&$ -3.378$&$ -2.985$\\
{\sc model9}&$-4.272$&$ -4.085$&$ -3.760$&$ -3.345$&$ -2.990$\\
{\sc model10}&$-4.347$&$ -4.138$&$ -3.803$&$ -3.362$&$ -2.978$\\
\hline\hline
\end{tabular}
\end{table}
\begin{table*}
\caption{Objects with the 5 highest reconstruction errors in the 10 trained models (The objects discussed in Section~\ref{sec:unsuccess} are highlighted by bold letters.)}
\centering
\label{table:errors}
\begin{tabular}{r|rr||r|rr}
model name&error&MaNGA ID&model name&error&MaNGA ID\\
\hline\hline
{\sc model1}
&0.002522&{\bf 60-1436778955512349056}&{\sc model6}&0.002936&{\bf 7-17219806}\\
&0.002360&{\bf 3-115120061}&&0.002429&{\bf 3-115120061}\\
&0.002081&3-105740024&&0.002314&{\bf 60-1436778955512349056}\\
&0.002024&7-14573163&&0.002135&60-3615860471750191104\\
&0.001957&3-151672729&&0.002056&3-151672729\\
\hline
{\sc model2}
&0.002529&{\bf 7-17219806}&{\sc model7}&0.002923&{\bf 3-115120061}\\
&0.002183&{\bf 3-115120061}&&0.002763&60-3615860471750191104\\
&0.002102&3-1516727296&&0.002747&{\bf 7-17219806}\\
&0.002093&60-3615860471750191104&&0.002528&3-151672729\\
&0.002052&3-105740024&&0.002133&7-14573163\\
\hline
{\sc model3}
&0.002289&{\bf 3-115120061}&{\sc model8}&0.002591&{\bf 3-115120061}\\
&0.002062&3-151672729&&0.002085&54-44051315\\
&0.002045&3-105740024&&0.002071&3-105740024\\
&0.002004&7-14573163&&0.002054&7-14573163\\
&0.001831&3-122926611&&0.001993&3-151672729\\
\hline
{\sc model4}
&0.002979&{\bf 3-33352569}&{\sc model9}&0.003321&{\bf 3-33352569}\\
&0.002317&{\bf 60-1436778955512349056}&&0.002200&{\bf 3-115120061}\\
&0.002230&{\bf 3-115120061}&&0.002186&3-151672729\\
&0.002113&3-105740024&&0.002145&7-14573163\\
&0.002088&7-14573163&&0.002075&3-105740024\\
\hline
{\sc model5}
&0.002120&{\bf 60-1436778955512349056}&{\sc model10}&0.002319&{\bf 3-33352569}\\
&0.002050&3-105740024&&0.002308&{\bf 3-115120061}\\
&0.002047&7-14573163&&0.002125&7-14573163\\
&0.002005&{\bf 3-115120061}&&0.002086&3-105740024\\
&0.001858&3-122926611&&0.002007&3-151672729\\
\hline\hline
\end{tabular}
\end{table*}

The near-continuous nature of the error distributions across almost the entire range indicates that the autoencoder models successfully capture the dominant patterns of common stellar spectra and accurately reproduce their features for the majority of the sample.
Figure~\ref{fig:A_star} shows a representative example of input spectra and their corresponding reconstructions produced by the trained autoencoder models.
The reconstructed spectra closely match the inputs, preserving the overall spectral shape and common features such as absorption lines.
In particular, prominent Balmer absorption lines, including H$\alpha$ ($\lambda = 6563\,\mathring{\mathrm{A}}$), H$\beta$ ($4861\,\mathring{\mathrm{A}}$), and H$\gamma$ ($4340\,\mathring{\mathrm{A}}$), as well as other metal lines, are reproduced with high fidelity.
This demonstrates that the autoencoder models successfully learn these absorption features as common characteristics of the training data.

A notable property of the reconstructed spectra is that they appear smoother and cleaner than the corresponding input spectra.
This behavior is characteristic of autoencoder-based approaches (\cite{V08,V10}). 
By compressing the input data into a low-dimensional latent space (Figure~\ref{fig:schematic}), the autoencoder is forced to learn only the most salient and recurring features of the input spectra.
As a result, uncorrelated noise features, whose shapes vary from measurement to measurement, are not retained in the latent representation.
The trained models therefore preferentially extract and reconstruct common, physically meaningful spectral features from the training sample.
Consequently, the residuals between the input and reconstructed spectra (shown in the upper panels of Figure~\ref{fig:A_star}) exhibit similar patterns, likely dominated by measurement noise and data-reduction artifacts.

One potential drawback of this behavior is that autoencoder models may also learn features arising from instrumental effects or telluric contamination if they are common among the input spectra.
Nevertheless, the successful reconstructions presented here demonstrate that the adopted autoencoder architecture is sufficiently expressive to capture the dominant physical features of the stellar spectra in the MaStar dataset.

This behavior supports the use of the reconstruction error as an effective anomaly score.
As discussed below, a subset of stellar spectra is not well reproduced by the trained models and appears in the high-error tail of the distributions shown in Figure~\ref{fig:error}.

\section{Unsuccessfully reconstructed cases}\label{sec:unsuccess}
\begin{figure*}
\begin{center}
\includegraphics[scale=0.70]{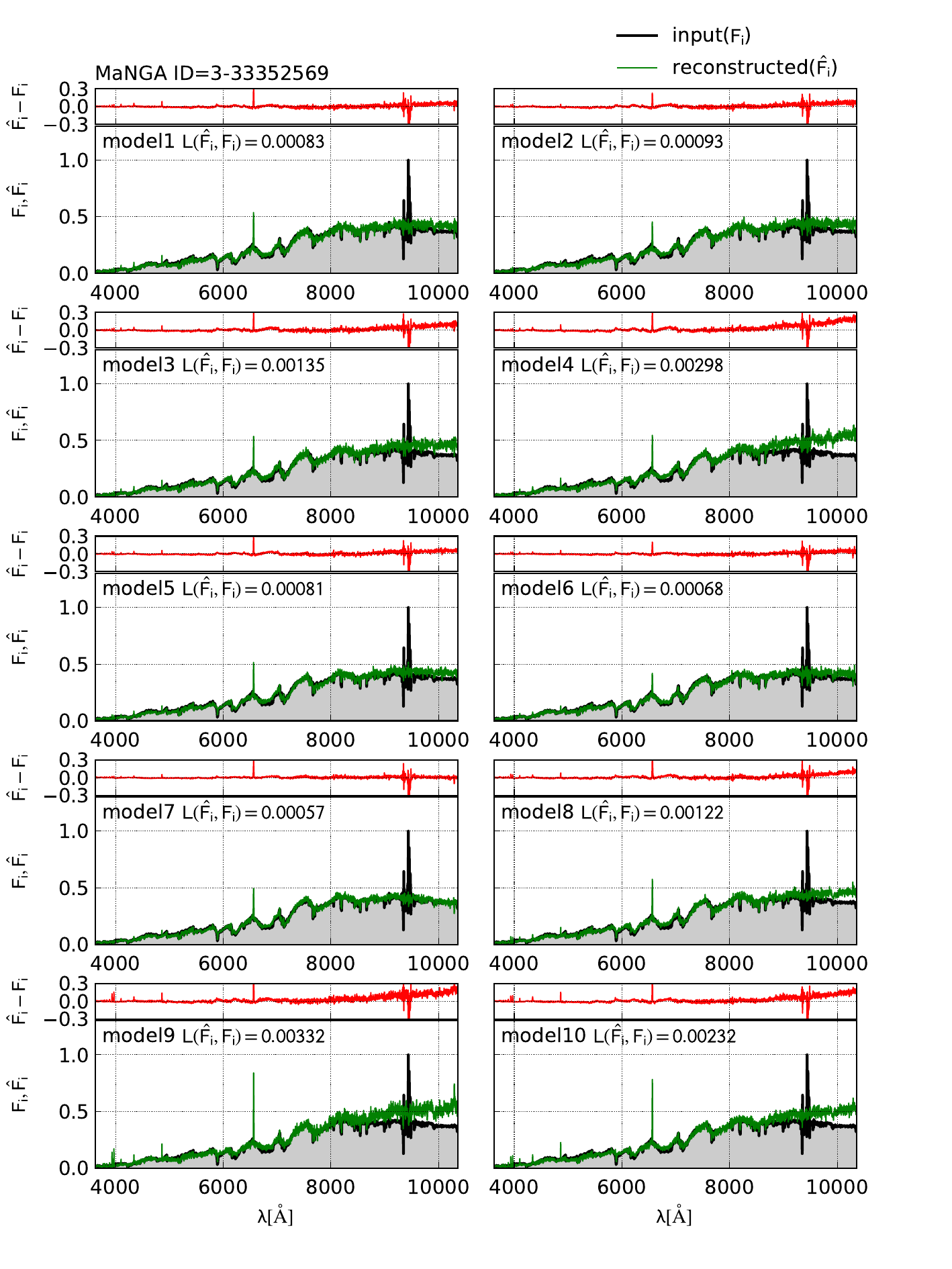}
\end{center}
\caption{Same as Figure~\ref{fig:A_star}, but for the unsuccessful case of the object specified by MaNGA ID 3-33352569. 
{Alt text: Ten plots showing the model and observed spectra wit thin and shick lines and residuapl plotted on top of each panel. }}
\label{fig:M_star}
\end{figure*}

In this section, we focus on cases of unsuccessful reconstruction and discuss possible sources of large reconstruction errors.
Table~\ref{table:errors} lists the five objects with the highest reconstruction errors for each of the ten trained autoencoder models.
Among these, four objects with MaNGA IDs \verb|7-17219806|, \verb|3-33352569|, \verb|3-115120061|, and \verb|60-1436778955512349056| appear frequently across the different models (highlighted in Table~\ref{table:errors}). 
These objects are found to be a spectrum affected by instrumental or reduction artifacts, two carbon stars, and an oxygen-rich TP-AGB star.
In the following, these three categories are examined in more detail.

\subsection{Object with possible artifacts}\label{sec:artifacts}
Figure~\ref{fig:M_star} presents the input and reconstructed spectra for the object with MaNGA ID \verb|3-33352569|, shown in the same format as Figure~\ref{fig:A_star}.
Based on visual inspection, the overall spectral shape is broadly consistent with that of an M-type star. 
Through its Gaia ID, I searched the counterpart source in SIMBAD Astronomical Database \footnote{https://simbad.cds.unistra.fr/simbad/}. 
Indeed, this object is turned out to be included in the APOGEE M-dwarf catalog (\cite{2017AJ....154...28B}) with an APOGEE ID of \verb|2M21140743+4924350| and estimated stellar parameters of effective temperature $T_\mathrm{eff} = 3428\,\mathrm{K}$ and metallicity $[\mathrm{M/H}] = -0.61$. 
These properties assures that this object is a commonly found M-type star. 

However, as shown in the figure, the input spectrum exhibits unusually high flux values around $\sim 9500\,\mathring{\mathrm{A}}$.
The presence of this feature causes several reconstructed spectra (notably {\sc model}4, {\sc model}9, and {\sc model}10) to significantly overestimate the flux in this wavelength region.
This localized overestimation leads to a large reconstruction error, resulting in this object being identified as one of the most poorly reconstructed cases.

The unsuccessful reconstruction suggests that the feature near $9500\,\mathring{\mathrm{A}}$ is not represented in the training dataset.
To our knowledge, however, there is no well-established stellar spectral feature expected at this wavelength for M-type stars.
If the excess flux were due to intrinsic stellar activity, such as emission associated with chromospheric processes, one might expect corresponding emission features in the Balmer lines (\cite{1974ApJS...28....1J,1979ApJ...234..579C}); however, no such emission is observed.
Therefore, it is concluded that the anomalous feature is most likely caused by instrumental effects or data-reduction artifacts rather than intrinsic stellar properties.

\begin{figure}
\begin{center}
\includegraphics[scale=0.50]{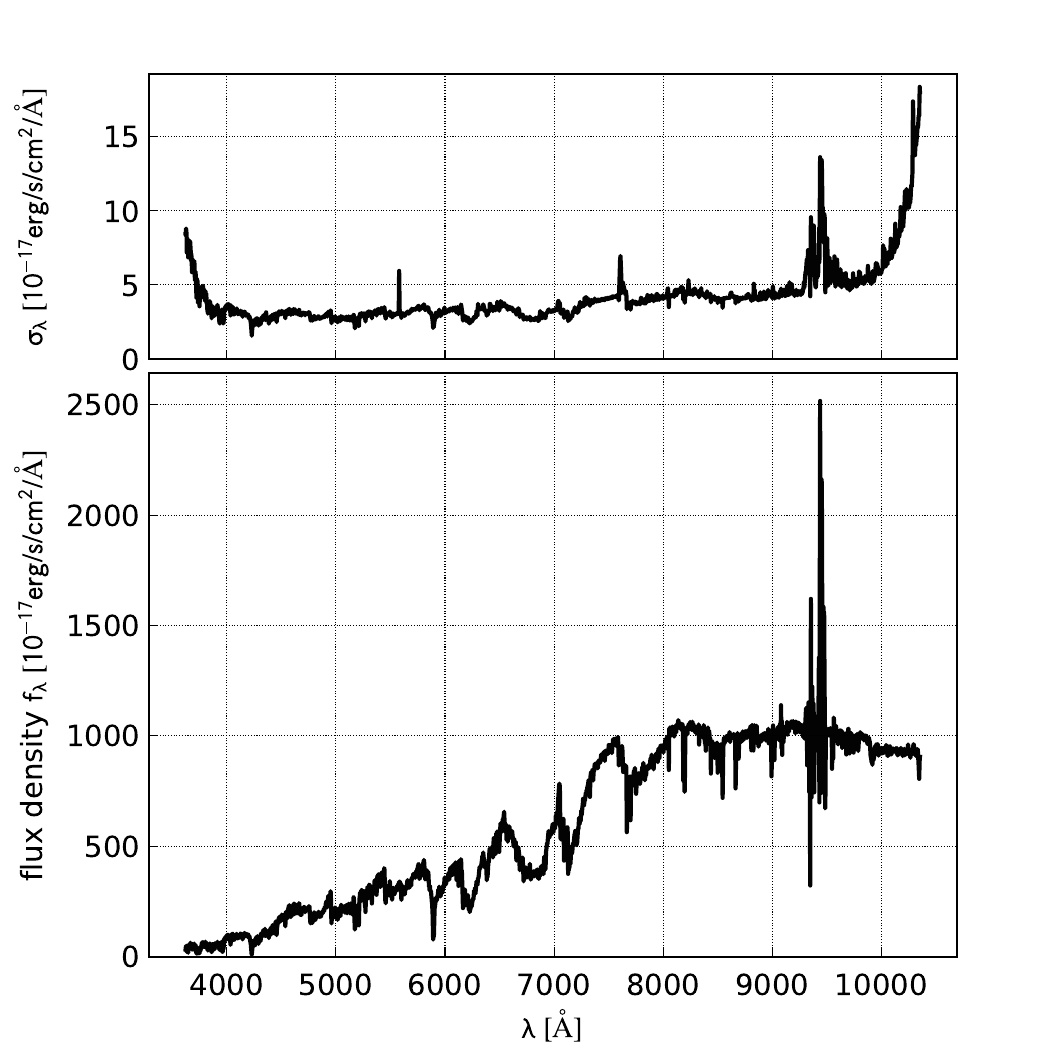}
\end{center}
\caption{Flux density per unit wavelength (lower) and the uncertainty $\sigma_\lambda$ (upper) of MaNGA ID 3-33352569. 
{Alt text: two line plots showing the spectrum and its uncertainty as a function of wavelength. }}
\label{fig:sigma_lambda}
\end{figure}

The MaSTAR \texttt{GOODVISIT} catalog parameter \texttt{BADPIXFRAC}, which quantifies the fraction of pixels with zero inverse variance, is equal to zero for this object, indicating that no pixels were fully masked by the reduction pipeline.
We therefore examined the inverse-variance ($\mathrm{IVAR}$) array provided with the MaStar spectra and converted it to per-pixel uncertainties via $\sigma_\lambda = 1/\sqrt{\mathrm{IVAR}}$.
Figure~\ref{fig:sigma_lambda} shows $\sigma_\lambda$ for this object together with the original spectrum.
Although no pixels are formally masked (\texttt{BADPIXFRAC} $= 0$), the region around $9500\,\mathring{\mathrm{A}}$ exhibits a localized increase in $\sigma_\lambda$ relative to the median uncertainty across the spectrum.
While the uncertainty continues to rise toward longer wavelengths, the anomaly detected by the trained models is confined to the enhancement near $9500\,\mathring{\mathrm{A}}$, suggesting a reduction artifact rather than an astrophysical feature. 

\subsection{Carbon star}\label{sec:carbon_star}

\begin{figure*}
\begin{center}
\includegraphics[scale=0.70]{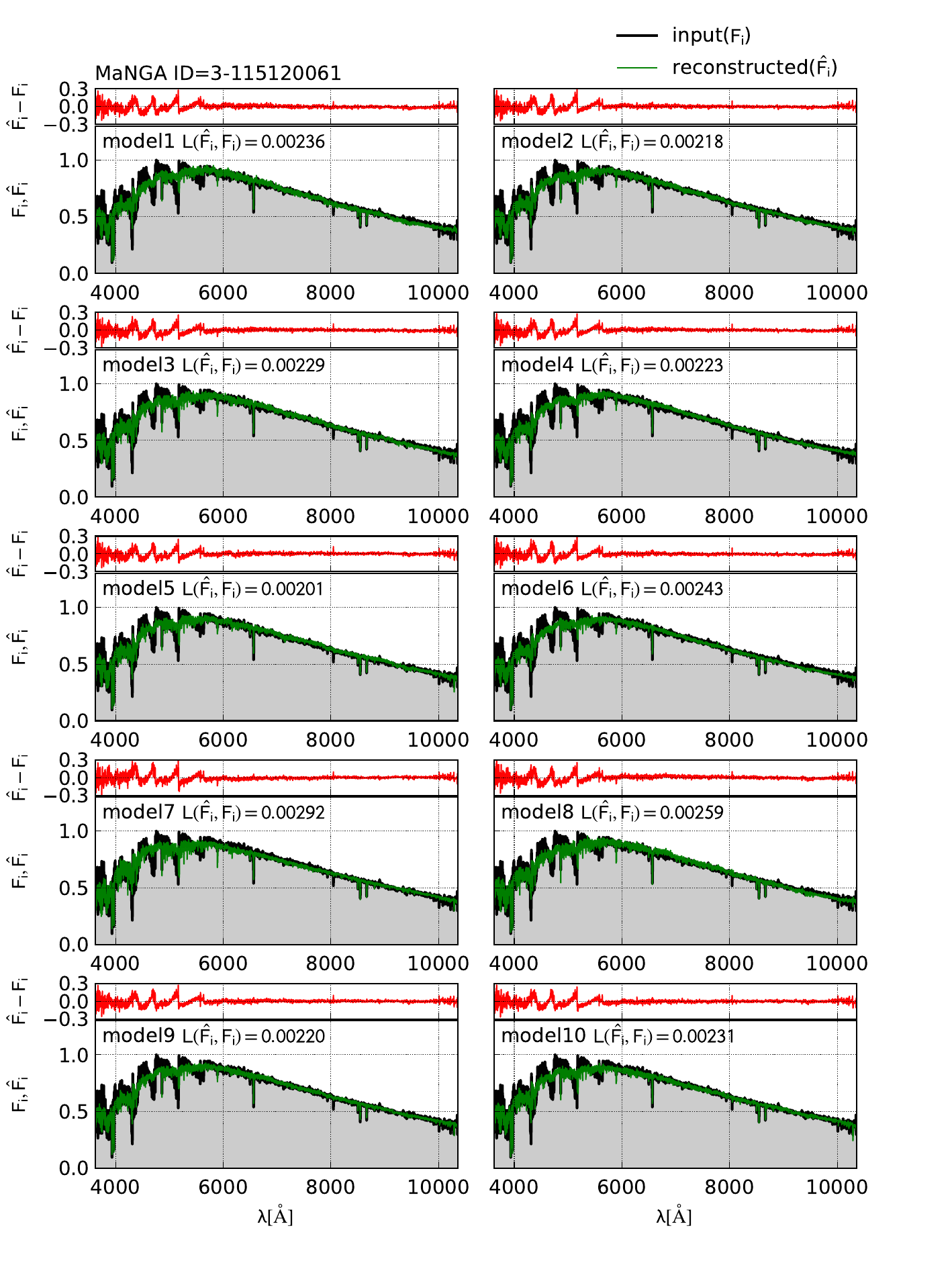}
\end{center}
\caption{Same as Figure~\ref{fig:A_star}, but for the unsuccessful case of the object specified by MaNGA ID 3-115120061. 
{Alt text: Ten plots showing the model and observed spectra wit thin and shick lines and residuapl plotted on top of each panel. }}
\label{fig:CH_star}
\end{figure*}
\begin{figure*}
\begin{center}
\includegraphics[scale=0.70]{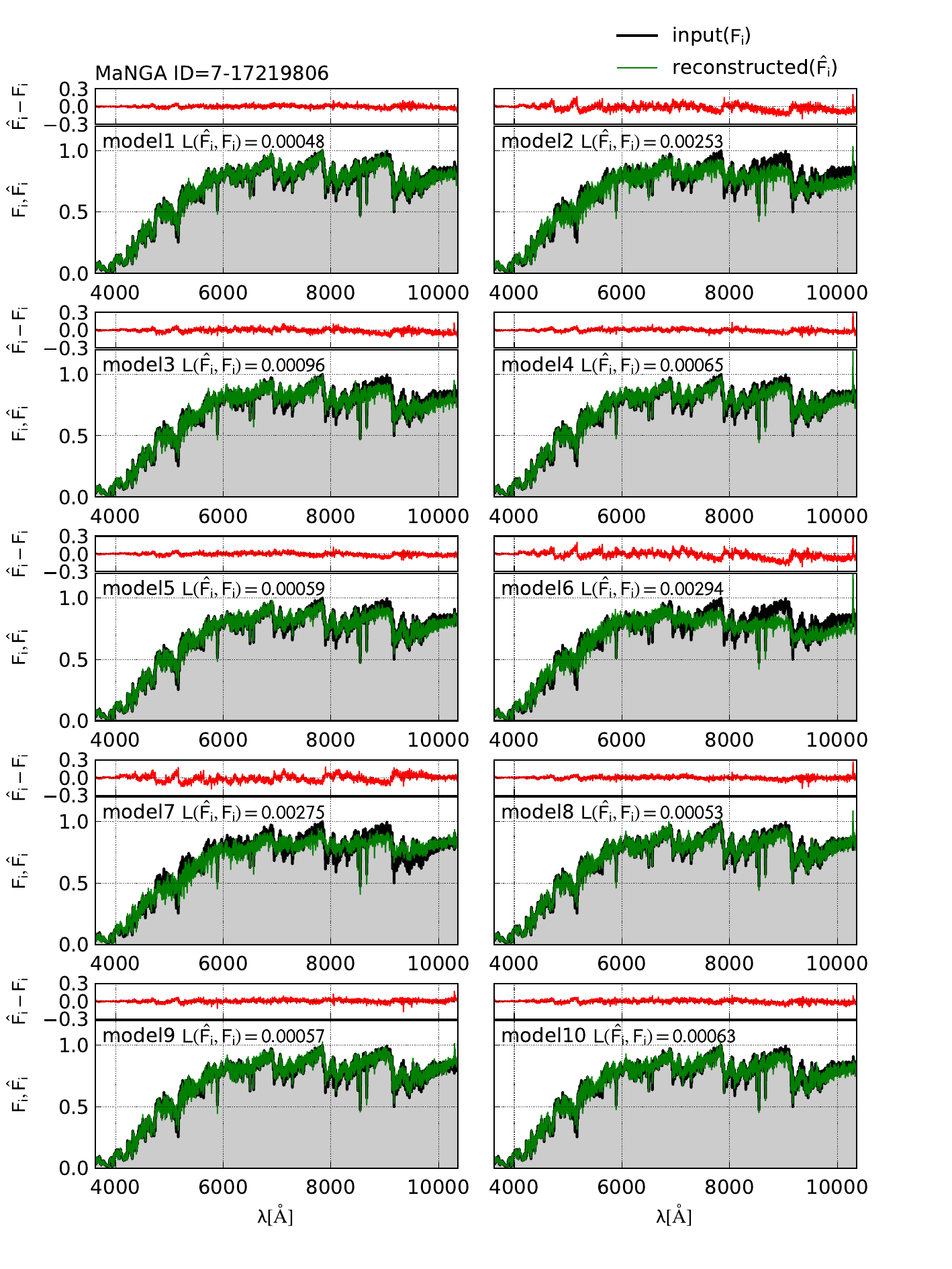}
\end{center}
\caption{Same as Figure~\ref{fig:A_star}, but for the unsuccessful case of the object specified by MaNGA ID MaNGA ID 7-17219806. 
{Alt text: Ten plots showing the model and observed spectra wit thin and shick lines and residuapl plotted on top of each panel. }}
\label{fig:C_star}
\end{figure*}
\begin{figure}
\begin{center}
\includegraphics[scale=0.7]{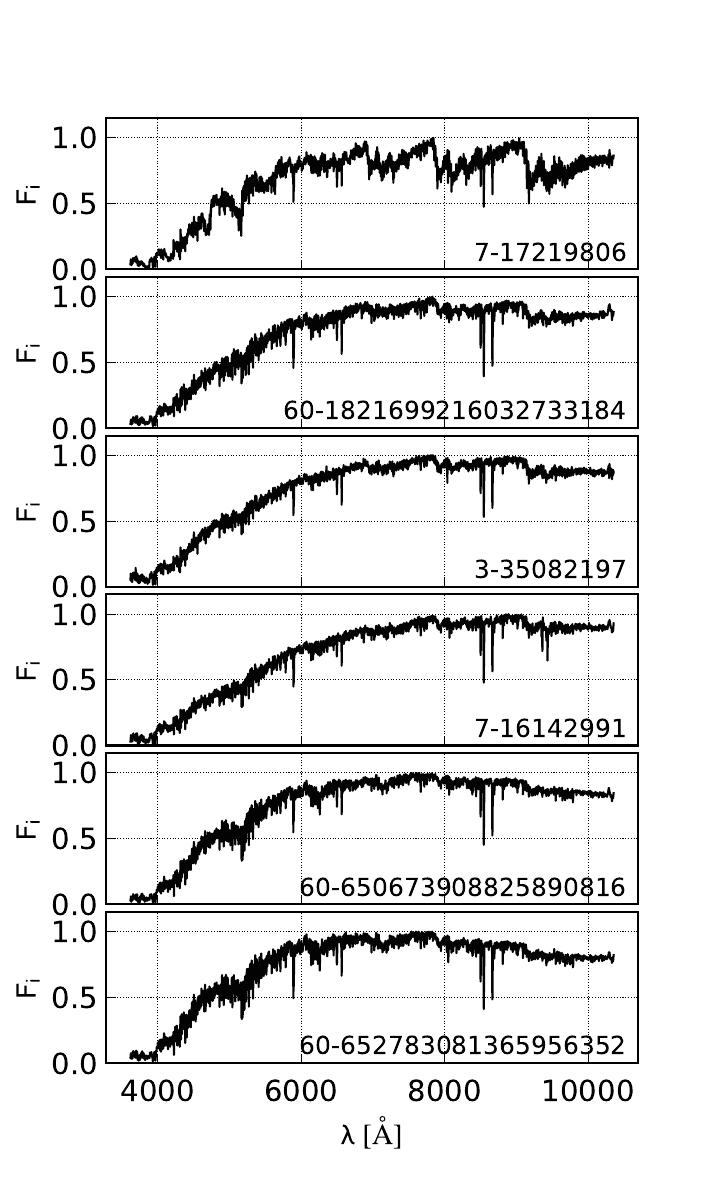}
\end{center}
\caption{Comparison of the spectrum of MaNGA ID MaNGA ID 7-17219806 (top) and neearest five objects in the latent space. 
{Alt text: Six line plots showing spectra of six objects in total. }}
\label{fig:similar_stars}
\end{figure}
\begin{figure}
\begin{center}
\includegraphics[scale=0.75]{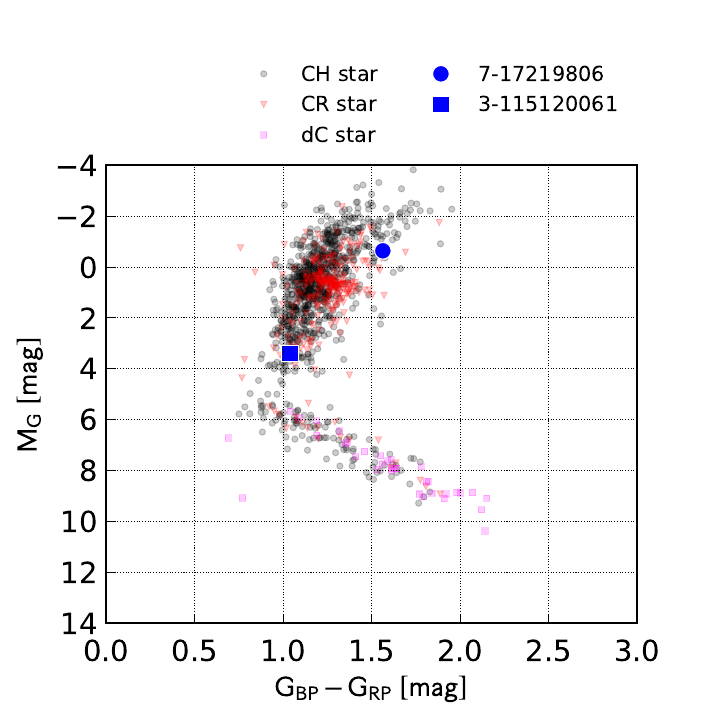}
\end{center}
\caption{Color-magnitude diagram fpr carbon stars in litterature (circle: CH star, lower triangle: CR star, and square: dC star). 
The two carbon stars flagged as outliers by the autoencode models are also presented (large square and circle). 
{Alt text: A scatter plot in the color and absolute magnitude plane showing carbon stars. }}
\label{fig:carbon_stars}
\end{figure}

We next turn to the second category of unsuccessful reconstructions, represented by the two objects with MaNGA IDs \verb|3-115120061| and \verb|7-17219806|.
Figures~\ref{fig:CH_star} and \ref{fig:C_star} present the input and reconstructed spectra for these two objects, respectively.
Unlike the previous example, the large reconstruction errors in these cases arise from intrinsic stellar spectral features rather than instrumental effects.

\subsubsection{Object with MaNGA ID 3-115120061}

The input spectrum shown in Figure~\ref{fig:CH_star} exhibits prominent spectral features in the wavelength range $4000$–$6000\,\mathring{\mathrm{A}}$ that deviate significantly from the typical stellar spectra represented in the training dataset.
As seen in the residuals, $\hat{\mathbf{F}} - \mathbf{F}$, shown in the upper panels of the figure, these features constitute the dominant source of the reconstruction error.
This behavior is consistently observed across all independently trained autoencoder models. 

The observed spectral characteristics are consistent with those of carbon stars (e.g., \cite{1993PASP..105..905K, 1998ARA&A..36..369W, 2016A&A...589A..36G}).
Carbon stars are generally characterized by strong molecular absorption features, most notably the CH G band near $4300\,\mathring{\mathrm{A}}$, $\mathrm{C}_2$ Swan bands at approximately $4300$–$6200\,\mathring{\mathrm{A}}$, and CN bands spanning $\sim 6000$–$10000\,\mathring{\mathrm{A}}$, along with additional molecular features.
These spectral signatures clearly distinguish carbon stars from the more common stellar spectra dominating the training dataset.
This object is indeed classified as a CH star in the latest LAMOST carbon star catalog (\cite{2025ApJS..279...40Z}; see also \cite{2015RAA....15.1671S, 2016ApJS..226....1J, 2018ApJS..234...31L, 2024ApJS..271...12L}).

\subsubsection{Object with MaNGA ID 7-17219806}

The object shown in Figure~\ref{fig:C_star} exhibits similar molecular band features, extending to longer wavelengths. 
In contrast to the CH star discussed above, several trained models (e.g., {\sc model1}, {\sc model8}, {\sc model9}, and {\sc model10}) are relatively successful in reproducing these features across the full wavelength range, resulting in moderate reconstruction errors.
This suggests that similar spectra are present in the training dataset, although their population may not be sufficiently large for the autoencoder to fully capture the diversity in feature strengths. 

Then, the training samples in {\sc model1}, which most successfully reproduced this object, are examined.
Objects located in close proximity to this source in the 10-dimensional latent space are identified.
Figure~\ref{fig:similar_stars} shows the spectra of the five nearest neighbors.
Because the deterministic autoencoder employed in this work does not impose any explicit regularization on the latent structure, the interpretation of distances in latent space is not straightforward.
Nevertheless, the objects shown in Figure~\ref{fig:similar_stars} display weaker, but similar spectral features to those of this object (MaNGA ID \texttt{7-17219806}), including moderate absorption edges across $7000$--$9000\,\mathring{\mathrm{A}}$.
Although the carbon features are significantly weaker than in the target object, the presence of such spectra within the training set may enable the autoencoder to learn the underlying structure and effectively extrapolate these features in more extreme cases.

This object corresponds to SDSS J190419.69+401520.5, which was previously identified as a carbon star by \citet{2013ApJ...765...12G}, who conducted a systematic survey of carbon stars in the SDSS.
More recently, it has been claimed to be a non-AGB or dwarf carbon star by \citet{2024MNRAS.530.1534H} (their Figure 10), who analyzed spectra from the MaStar datasets by combining color selection and spectral fitting.

\subsubsection{Properties of two carbon stars}
For further assessment of the identities of these two objects, Their absolute magnitudes and BP–RP colors are compared with those of known carbon star populations spanning different subclasses. 
I used samples of two subtypes of classical carbon stars (CH and CR stars) from \citet{2018ApJS..234...31L} and dwarf carbon stars (dC) from \citet{2021ApJ...922...33R}. Figure~\ref{fig:carbon_stars} shows the resulting color–magnitude diagram.

The carbon star population exhibits two distinct branches: a bright branch and a faint branch. Dwarf carbon stars are main-sequence stars with enhanced surface carbon abundances \citep{1977ApJ...216..757D} likely due to binary mass transfer and therefore predominantly occupy the faint branch. In contrast, classical carbon stars classified as CH and CR stars are more evolved and are mainly found along the bright branch.

As shown in Figure~\ref{fig:carbon_stars}, the two objects identified as anomalous are located on the bright branch. For the object with MaNGA ID \verb|3-115120061|, this provides further confirmation of its consistency with a CH star classification. The other object, with MaNGA ID \verb|7-17219806|, is located in an even brighter region of the diagram. This placement favors a classification as a classical carbon star rather than a dwarf carbon star.

In summary, the two objects flagged as anomalous are carbon stars exhibiting characteristic spectral features. 
Their positions in the color–magnitude diagram (Figure~\ref{fig:cmd}) indicate that they are not isolated from the broader stellar population in the dataset. 
This suggests that the stellar parameters shaping their continua, such as effective temperature and surface gravity, are not unusual. 
However, the enhanced surface carbon abundance, and the resulting presence of strong molecular bands (C$_2$, CN, and CH), ultimately lead to unsuccessful spectral reconstructions, as highlighted in Figures~\ref{fig:CH_star} and \ref{fig:C_star}.

\subsection{Oxygen-rich thermally pulsating AGB star}\label{sec:late_M_star}
\begin{figure*}
\begin{center}
\includegraphics[scale=0.70]{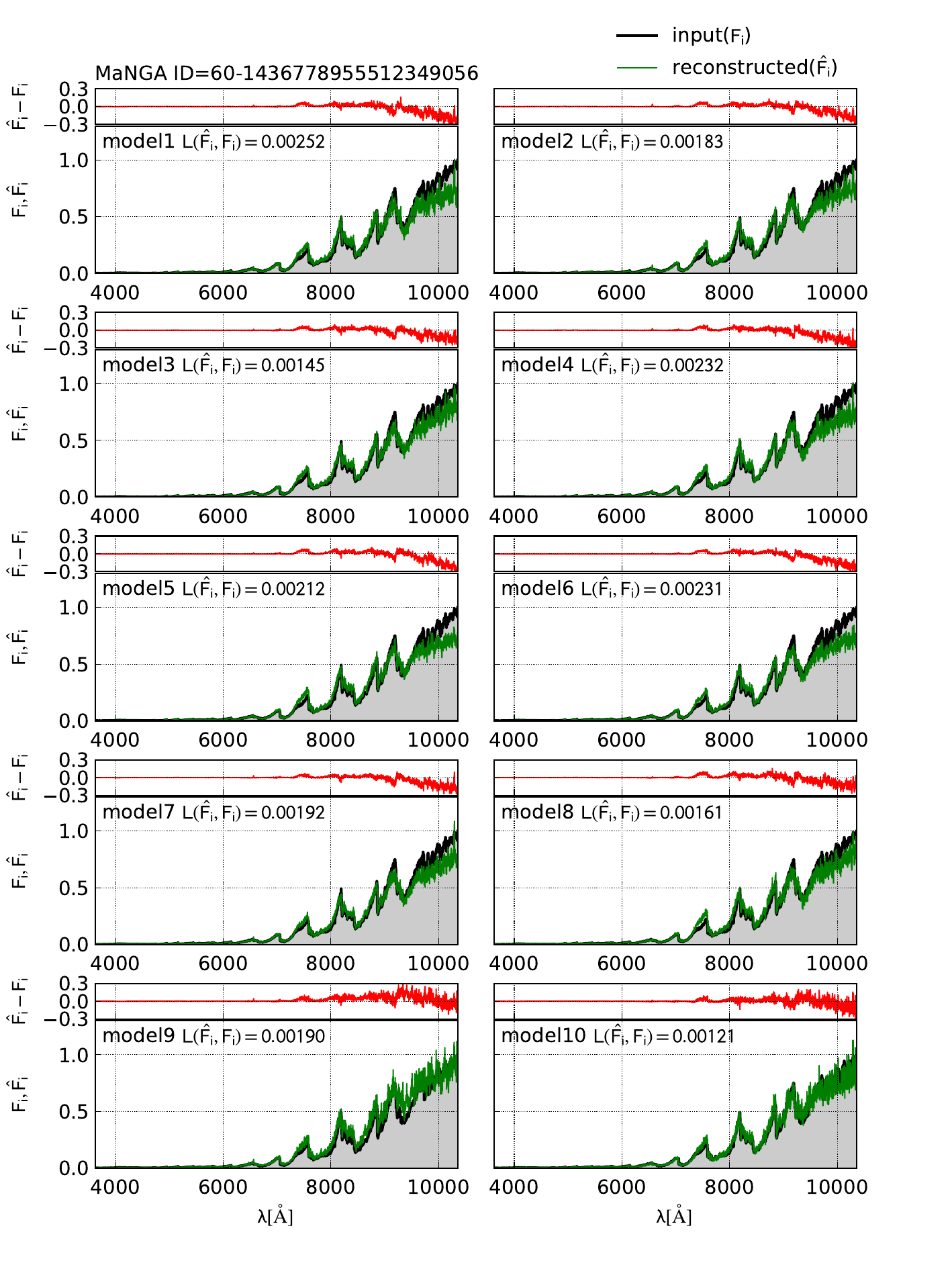}
\end{center}
\caption{Same as Figure~\ref{fig:A_star}, but for the unsuccessful case of the object specified by MaNGA ID MaNGA ID 0-1436778955512349056. 
{Alt text: Ten plots showing the model and observed spectra wit thin and shick lines and residuapl plotted on top of each panel. }}
\label{fig:late_M_star}
\end{figure*}
\begin{figure}
\begin{center}
\includegraphics[scale=0.70]{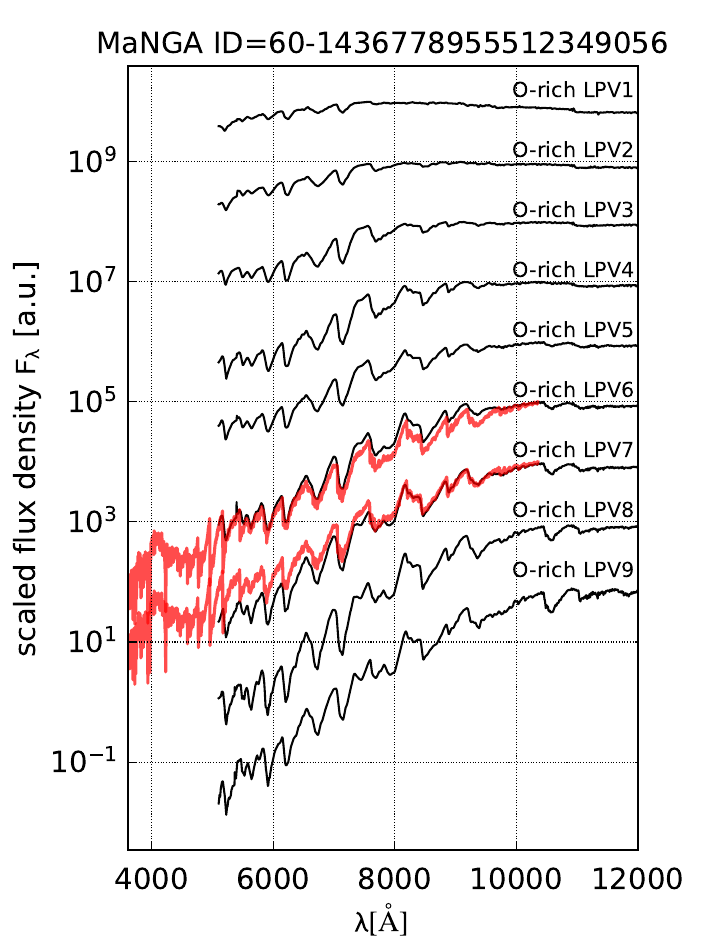}
\end{center}
\caption{Spectral comparison of the object with MaNGA ID 60-1436778955512349056 (thick line) and oxygen-rich LPVs (thin lines). The nine mean spectra of O-rich LPVs are ploted from top to bottom in order of their colors (bluer to redder). 
{Alt text: Plot showing nine different template spectra and two spectra from the data. }}
\label{fig:o_rich_agb}
\end{figure}

Finally, Figure~\ref{fig:late_M_star} presents the case of the object with MaNGA ID \verb|60-1436778955512349056|.
The input spectrum is characterized by an extremely red color, with almost negligible flux shortward of $\sim6000\,\mathring{\mathrm{A}}$.
The trained models systematically underestimate the flux at the longest wavelengths ($\gtrsim 9500\,\mathring{\mathrm{A}}$), which serves as the dominant source of the large reconstruction error. 
Such red stars are intrinsically faint in the optical and are therefore underrepresented in spectroscopic surveys. 
Despite their abundance in the Galaxy, their sparsity in the training dataset can generally lead to relatively higher reconstruction errors when analyzed with autoencoder models trained predominantly on earlier-type stars. 

The extremely red nature of this object is highlighted in the color–magnitude diagram shown in Figure~\ref{fig:cmd}.
With $\mathrm{M}_G = -0.19$ and $\mathrm{G}_{BP}-\mathrm{G}_{RP}= 4.46$, this object is one of the reddest sources in the entire dataset, located at the end of the asymptotic giant branch. 
As shown in Figure~\ref{fig:cmd}, the region with BP-RP color esceeding $4.0$ is only sparsely populated, containing fewer than 10 objects. 
This scarcity forces the autoencoder models to effectively extrapolate spectra during reconstruction.
This case therefore suggests that autoencoder-based anomaly detection can also flag samples at the extrema of continuous physical distributions. 

The region around this object in the color–magnitude diagram (Figure~\ref{fig:cmd}) is known to be occupied by TP-AGB stars, which generally exhibit long-period variability.
Consistently, this object, also known as IRAS~16572+5843, was identified in the Northern Sky Variability Survey \citep{2004AJ....128.2965W} as a long-period variable (LPV) with a period of $730$ days.
\citet{2000A&AS..146..217L} published a spectral library of luminous cool stars, including LPVs, covering the wavelength range $5000$--$25000\,\mathring{\mathrm{A}}$.
Subsequently, \citet{2002A&A...393..167L} constructed averaged spectra of carbon(C)- and oxygen(O)-rich LPVs by grouping samples according to their colors.
In Figure~\ref{fig:o_rich_agb}, the spectrum of this object is compared with the nine mean spectra of O-rich LPVs from \citet{2002A&A...393..167L} (labeled as O-rich LPV1 through LPV9, ordered by increasing color).
Based on visual inspection, this object is most consistent with O-rich LPV6 or LPV7 as shown in the figure. 
Although a more quantitative comparison is beyond the scope of this study, the notably good agreement supports the identification of this object as an O-rich TP-AGB star.
TP-AGB stars originate from $\sim1$--$8\,M_\odot$ stars and represent a short-lived evolutionary phase lasting up to a few $10^6\,\mathrm{yr}$.
In this sense, the autoencoder models successfully identify an object in a rare evolutionary stage. 

\section{Discussion and conclusion}\label{sec:summary}
The results of this study show that autoencoder-based models can efficiently identify both non-physical anomalies (Sec.~\ref{sec:artifacts}) and astrophysically interesting rare objects (Sec.\ref{sec:carbon_star} and \ref{sec:late_M_star}) without relying on prior assumptions about stellar classification. 
In the former case, such detections can be leveraged to identify spectral regions that should be masked or removed in subsequent analyses. 
In the latter case, the method enables the discovery of objects with unusual spectral features (Section~\ref{sec:carbon_star}) as well as sources occupying the extremes of continuous physical distributions (Section~\ref{sec:late_M_star}). 
The models therefore provides a robust framework for exploring large spectroscopic surveys.

While autoencoder-based anomaly detection has been increasingly applied to spectroscopic datasets, identifying the sources of reconstruction error is essential not only for a deeper understanding of the method, but also for guiding its future extensions (e.g., \cite{2025A&A...703A.242O}). 
This work showcases examples of unsuccessful reconstructions from trained autoencoder models applied to the MaNGA Stellar Library, combining reconstruction-error analysis with careful examination of spectral morphology and data-quality indicators. 
The focus is on clarifying the types of objects flagged as anomalous and helps distinguish between astrophysical peculiarities and other sources of deviation.

Future studies could extend autoencoder-based anomaly detection to larger and more diverse stellar spectral datasets, including ongoing and upcoming surveys such as SDSS-V (\cite{2017arXiv171103234K}), WEAVE (\cite{2024MNRAS.530.2688J}), 4MOST (\cite{2019Msngr.175....3D}), and the Prime Focus Spectrograph (PFS) on {\it Subaru} (\cite{2014PASJ...66R...1T}). Incorporating multi-epoch or multi-wavelength data could help capture variable or rare spectral features.
Furthermore, systematic follow-up of objects flagged as anomalous offers opportunities for revealing rare spectral types, chemically peculiar stars, or unusual evolutionary states. 
This is essential to maximize the scientific return of large spectroscopic surveys.

Methodological improvements, such as variational autoencoders (\cite{2013arXiv1312.6114K, Rezende2014}) and attention-based architectures (\cite{2017arXiv170603762V}), may enhance the model’s ability to capture complex correlations and better distinguish subtle anomalies from noise. Integrating stellar parameters, including effective temperature, surface gravity, and metallicity, as auxiliary inputs could further improve anomaly detection conditional on spectral type.

Overall, this study highlights the potential of unsupervised deep learning techniques to uncover unusual stellar spectra in large datasets, providing a valuable tool for both quality control and the discovery of astrophysically interesting objects.

\begin{ack}
The author thanks the anonymous referee for his/her constructive suggestions on the manuscript. 
The author appreciates Yohei Masada, Jin Matsumoto, and participants of Astroinformatics Advanced Programming Techniques (AAPT) workshop 2025 held at Fukuoka University, for fruitful discussion. 
The computations are carried out by using resceubbc PC cluster at the Research Center for the Early Universe, the Univsersity of Tokyo. 

This work made use of data from the MaNGA Stellar Library (MaStar) as part of the Sloan Digital Sky Survey IV (SDSS-IV). 
Funding for the Sloan Digital Sky 
Survey IV has been provided by the 
Alfred P. Sloan Foundation, the U.S. 
Department of Energy Office of 
Science, and the Participating 
Institutions. 

SDSS-IV acknowledges support and 
resources from the Center for High 
Performance Computing  at the 
University of Utah. The SDSS 
website is www.sdss4.org.

SDSS-IV is managed by the 
Astrophysical Research Consortium 
for the Participating Institutions 
of the SDSS Collaboration including 
the Brazilian Participation Group, 
the Carnegie Institution for Science, 
Carnegie Mellon University, Center for 
Astrophysics | Harvard \& 
Smithsonian, the Chilean Participation 
Group, the French Participation Group, 
Instituto de Astrof\'isica de 
Canarias, The Johns Hopkins 
University, Kavli Institute for the 
Physics and Mathematics of the 
Universe (IPMU) / University of 
Tokyo, the Korean Participation Group, 
Lawrence Berkeley National Laboratory, 
Leibniz Institut f\"ur Astrophysik 
Potsdam (AIP),  Max-Planck-Institut 
f\"ur Astronomie (MPIA Heidelberg), 
Max-Planck-Institut f\"ur 
Astrophysik (MPA Garching), 
Max-Planck-Institut f\"ur 
Extraterrestrische Physik (MPE), 
National Astronomical Observatories of 
China, New Mexico State University, 
New York University, University of 
Notre Dame, Observat\'ario 
Nacional / MCTI, The Ohio State 
University, Pennsylvania State 
University, Shanghai 
Astronomical Observatory, United 
Kingdom Participation Group, 
Universidad Nacional Aut\'onoma 
de M\'exico, University of Arizona, 
University of Colorado Boulder, 
University of Oxford, University of 
Portsmouth, University of Utah, 
University of Virginia, University 
of Washington, University of 
Wisconsin, Vanderbilt University, 
and Yale University.

This work has made use of data from the European Space Agency (ESA) mission Gaia (https://www.cosmos.esa.int/gaia), processed by the Gaia Data Processing and Analysis Consortium (DPAC). Funding for the DPAC has been provided by national institutions, in particular the institutions participating in the Gaia Multilateral Agreement.
\end{ack}

\section*{Funding}

\section*{Data availability} 
The trained models and parameters are available upon a reasonable request. 

\appendix 


\end{document}